\documentstyle[amsmath,amssymb,graphicx]{article}

\def\be{\begin{eqnarray}}
\def\ee{\end{eqnarray}}
\def\nn{\nonumber}

\def\tr{{\rm tr}\,}
\def\Tr{{\rm Tr}\,}

\def\l[{\phantom.[}

\hoffset= - 1.0in         % switch off for draft style
\title{{\bf Colored HOMFLY polynomials as multiple sums
over paths or standard Young tableaux} \vspace{.5cm}}
\author{{\bf A.Anokhina}\footnote{ {\small {\it
MIPT, Dolgoprudny, Russia} and {\it ITEP, Moscow, Russia}};
anokhina@itep.ru}, \ {\bf A.Mironov}\footnote{ {\small {\it Lebedev
Physics Institute} and {\it ITEP, Moscow, Russia}}; mironov@itep.ru;
mironov@lpi.ru}, \ {\bf A.Morozov}\thanks{{\small {\it ITEP, Moscow,
Russia}}; morozov@itep.ru}, \ {\bf And.Morozov}\thanks{{\small {\it
Moscow State University} and {\it ITEP, Moscow, Russia}};
Andrey.Morozov@itep.ru}\date{ }}

\begin{document}

\setcounter{footnote}{3}

\setcounter{tocdepth}{3}

\maketitle

\vspace{-5.5cm}

\begin{center}
\hfill FIAN/TD-06/13\\
\hfill ITEP/TH-13/13
\end{center}

\vspace{4cm}

\begin{abstract}
If a knot is represented by an $m$-strand braid,
then HOMFLY polynomial in representation $R$
is a sum over characters in all representations
$Q\in R^{\otimes m}$.
Coefficients in this sum are traces of products
of quantum $\hat{\cal R}$-matrices along the braid,
but these matrices act in the space of intertwiners,
and their size is equal to the multiplicity $M_{RQ}$ of $Q$
in $R^{\otimes m}$.
If $R$ is the fundamental representation $R=[1]=\Box$, then
$M_{\Box Q}$ is equal to the number of paths
in representation graph, which lead from the
fundamental vertex $\Box$ to the vertex $Q$.
In the basis of paths the entries of the $m-1$ relevant
$\hat{\cal R}$-matrices
are associated with the pairs of paths and
are non-vanishing only when the two paths either coincide
or differ by at most one vertex;
as a corollary
$\hat{\cal R}$-matrices consist of just $1\times 1$ and
$2\times 2$ blocks, given by very simple explicit expressions.
If cabling method is used to color the knot with the
representation $R$, then the braid has $m|R|$ strands,
$Q$ have a bigger size $m|R|$,
but only paths passing through the vertex $R$ are
included into the sums over paths which define the
products and traces of the $m|R|-1$ relevant $\hat{\cal R}$-matrices.
In the case of $SU(N)$
this path sum formula can also be interpreted as
a multiple sum over the standard Young tableaux.
By now it
provides the most effective way for evaluation
of the colored HOMFLY polynomials,
conventional or extended,
for arbitrary braids.
\end{abstract}

\bigskip

\bigskip

\section{Introduction}

Knot polynomials are currently among the central objects of interest
in quantum field theory: they are exactly at the border between
the known and unknown.
The knot polynomials can be defined as Wilson loop averages
\be
H_R^{\cal K\subset {\cal M}}(q|G)
= \left< {\rm Tr}_R\ P\exp \left(\oint_{\cal K}  {\cal A}\right)\right>_{CS}
\ee
i.e. generic gauge-invariant observables in the simplest version of the
$3$-dimensional Yang-Mills theory,
the topological Chern-Simons model \cite{CS}
with the action
\be
\frac{k}{4\pi}\int_{\cal M} \Tr \Big({\cal A}d{\cal A} + \frac{2}{3}{\cal A}^3\Big)
\ee
and they depend on a closed contour ${\cal K}$ in a three dimensional
manifold ${\cal M}$, on the representation $R$ of the gauge group $G=SU(N)$
and on the coupling constant $q=e^{2\pi i/(k+N)}$.
Since the theory is topological, the dependence is actually only on the topological
class of ${\cal K}$, i.e. the contour can be considered as a knot.
For the simply connected space ${\cal M}=R^3$ or $S^3$ the average $H(q|G)$
is actually a polynomial in $q$ and $A=q^N$, hence, the name "knot polynomial".

The study of knot polynomials in topology goes back to
\cite{knots}, and they were put into the context of quantum field theory
in the seminal works by A.Schwarz \cite{Sch} and E.Witten \cite{Wit},
further developed in \cite{TR,AdvCS}.
Since the theory is topological, there are no dynamical phenomena like confinement,
instead a close relation exists to the $2$-dimensional conformal theories
\cite{Wit,CSth}, very much in the spirit of AdS/CFT correspondence \cite{AdS/CFT}.
In this way the knot polynomials are related to the most difficult part
of conformal field theory, to modular transformations,
which through the AGT relations \cite{AGT} are connected to the $S$-duality
between the ${\cal N}=2$ supersymmetric Yang-Mills theories \cite{Sdual}.
This makes the study of knot polynomials the next task after the structure
of conformal blocks themselves is more or less understood in terms
of the Dotsenko-Fateev matrix models \cite{DFcb} and other similar representations
\cite{Bel}.

An additional interest is induced by existence of non-trivial deformations
of the knot polynomials:
to superpolynomials \cite{KhR,sup,DMMSS}
and refined Chern-Simons theory \cite{refi}  (see also \cite{betadefo}),
to {\it extended} knot polynomials \cite{MMMkn1}, which
puts them into the class of $\tau$-function like objects, etc.
At least naively \cite{MMSle}, they belong to the family of Hurwitz
partition functions \cite{Hurw}, more general than the conventional
KP/Toda $\tau$-functions, probably related to the generalized
$\tau$-functions of \cite{GKLMM}.

However, for any kind of generic investigation and application
of knot polynomials, they should be first effectively calculated
and represented in a theoretically appealing form,
allowing evaluation of these polynomials for particular knots
and representations.
There are different competitive approaches to do this,
e.g. \cite{katlas,inds}.
The goal of this letter is to summarize the results of
our method \cite{MMMkn1,MMMkn2,AnoMMM,AnoM},
which provides a complete, nice
and practically efficient solution to this problem.

\section{HOMFLY polynomials via quantum ${\cal R}$-matrices}

The method may begin with choosing the temporal gauge $A_0=0$ \cite{MS}
in Chern-Simons theory, then the theory becomes quadratic with
the ultralocal propagator $\theta(t)\delta(\vec x)$.
Then the original knot in $3$-dimensions is substituted by a $2$-dimensional
knot diagram (a 4-valent oriented graph), and the Wilson average reduces to a
$q$-graded
trace of the product of quantum ${\cal R}$-matrices,
standing at the vertices of the graph \cite{TR}.

It is most convenient to choose the knot diagram in the form of
a closure of a braid. If the braid has $m$ strands, then the product
involves $m-1$ different ${\cal R}$-matrices:
${\cal R}_{(i)}$ stands at the intersection of strands $i$ and $i+1$,
and $i=1,\ldots,m-1$. For instance, for the $3$-strand braid
one has
\be
H_R^{(a_1,b_1|a_2,b_2|,\ldots)} = \Tr^{\rm grad}_{R^{\otimes 3}}\ {\cal R}_{(1)}^{a_1}
{\cal R}_{(2)}^{b_1} {\cal R}_{(1)}^{a_2} {\cal R}_{(2)}^{b_2} \ldots
\label{3str}
\ee
In the pattern picture $a_1=0,b_1= -2,a_2=2,b_2=-1,a_3=3$:

\bigskip

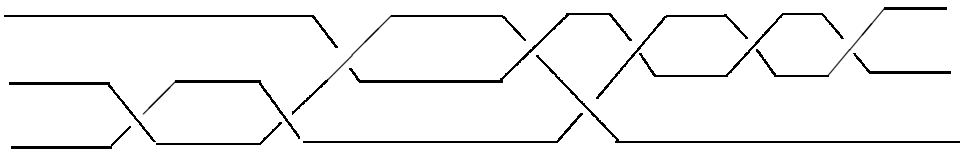
\unitlength 1mm % = 2.845pt
\linethickness{0.4pt}
\ifx\plotpoint\undefined\newsavebox{\plotpoint}\fi % GNUPLOT compatibility
\begin{picture}(145.5,53)(0,0)
\put(19.5,34.5){\line(1,0){13.25}}
\put(41.25,43.25){\line(1,0){11.25}}
\put(19.25,43){\line(1,0){13.25}}
\put(38.75,35){\line(1,0){13.75}}
\put(61.25,43.25){\line(1,1){8.75}}
\put(70,52){\line(1,0){14.75}}
\put(18.5,52){\line(1,0){41}}
%\emline(59.5,52)(62.75,47.75)
\multiput(59.5,52)(.033505155,-.043814433){97}{\line(0,-1){.043814433}}
%\end
\put(58.25,35.25){\line(1,0){33.75}}
%\emline(92,35.25)(95.25,39)
\multiput(92,35.25)(.033505155,.038659794){97}{\line(0,1){.038659794}}
%\end
%\emline(64.5,45)(65.75,43.25)
\multiput(64.5,45)(.03289474,-.04605263){38}{\line(0,-1){.04605263}}
%\end
\put(65.75,43.25){\line(1,0){19}}
%\emline(84.5,43.5)(93.5,52.25)
\multiput(84.5,43.5)(.0346153846,.0336538462){260}{\line(1,0){.0346153846}}
%\end
%\emline(84.75,52)(87.75,48.75)
\multiput(84.75,52)(.03370787,-.03651685){89}{\line(0,-1){.03651685}}
%\end
%\emline(52.5,43)(57.75,35.75)
\multiput(52.5,43)(.033653846,-.046474359){156}{\line(0,-1){.046474359}}
%\end
%\emline(52.5,35)(55.25,37.75)
\multiput(52.5,35)(.03353659,.03353659){82}{\line(0,1){.03353659}}
%\end
%\emline(56.75,39)(61.5,43.5)
\multiput(56.75,39)(.035447761,.03358209){134}{\line(1,0){.035447761}}
%\end
%\emline(32.25,43)(38.5,35.25)
\multiput(32.25,43)(.033602151,-.041666667){186}{\line(0,-1){.041666667}}
%\end
%\emline(32.75,34.75)(35.25,37.25)
\multiput(32.75,34.75)(.03333333,.03333333){75}{\line(0,1){.03333333}}
%\end
\put(37,39){\line(1,1){4.25}}
\put(99.75,35.25){\line(1,0){45.75}}
%\emline(100,35.5)(89.25,46.75)
\multiput(100,35.5)(-.0336990596,.0352664577){319}{\line(0,1){.0352664577}}
%\end
%\emline(97.25,41)(106.5,52)
\multiput(97.25,41)(.0336363636,.04){275}{\line(0,1){.04}}
%\end
\put(106.5,52){\line(1,0){7.75}}
\put(121.25,44){\line(1,0){6.75}}
\put(128,44){\line(5,6){7.5}}
\put(135.5,53){\line(1,0){8.25}}
\put(93.25,52.25){\line(1,0){5.75}}
%\emline(99,52.25)(101.75,48.75)
\multiput(99,52.25)(.03353659,-.04268293){82}{\line(0,-1){.04268293}}
%\end
%\emline(103,47)(105,44)
\multiput(103,47)(.03333333,-.05){60}{\line(0,-1){.05}}
%\end
\put(105,44){\line(0,1){0}}
\put(105,44){\line(1,0){9.5}}
%\emline(114.5,44)(122,52.25)
\multiput(114.5,44)(.033632287,.036995516){223}{\line(0,1){.036995516}}
%\end
\put(122,52.25){\line(1,0){5.25}}
%\emline(127.25,52.25)(130,49)
\multiput(127.25,52.25)(.03353659,-.03963415){82}{\line(0,-1){.03963415}}
%\end
%\emline(131.5,47)(133.5,44.5)
\multiput(131.5,47)(.03333333,-.04166667){60}{\line(0,-1){.04166667}}
%\end
\put(133.5,44.5){\line(1,0){10.75}}
%\emline(114.25,52.25)(117.25,49)
\multiput(114.25,52.25)(.03370787,-.03651685){89}{\line(0,-1){.03651685}}
%\end
%\emline(121,44)(118.5,47.5)
\multiput(121,44)(-.03333333,.04666667){75}{\line(0,1){.04666667}}
%\end
\end{picture}

\vspace{-2.7cm}

\noindent
Similarly, for arbitrary $m$
\be
H_R^{(a_{11},\ldots,a_{1,m-1}|a_{21},\ldots,a_{2,m-1}|,\ldots)}
= \Tr^{\rm grad}_{R^{\otimes m}}\ {\cal R}_{(1)}^{a_{11}}\ldots
{\cal R}_{(m-1)}^{a_{1,m-1}} {\cal R}_{(1)}^{a_{21}}\ldots  {\cal R}_{(m-1)}^{a_{2,m-1}} \ldots
\label{mstr}
\ee
The trace here is defined with additionally inserted element $q^{R^{\otimes m}}$ so that
\be
\chi_Q = \Tr^{\rm grad}_Q I ={\rm dim}_q^G(Q)
\ee
are the quantum dimensions of the representation $R$ (the characters of the group $G$ at the special values
$p_k=\frac{A^k-A^{-k}}{q^k-q^{-k}}$).
In this formula the ${\cal R}$ matrices are of the huge size ${\rm dim}(R)^2 \times {\rm dim}(R)^2$
and this expression can seem absolutely hopeless to evaluate for generic group $G$ and
representation $R$.

However, things are actually much more simple.
The product $R^{\otimes m}$ can be expanded into a sum of irreducible representations,
generically, with non-trivial multiplicities.
The crucial observation is that ${\cal R}_{(i)}$ act as unity in each irreducible representation,
so that the matrices in  (\ref{mstr})
can be reduced to $\hat{\cal R}_{(i)}$
of a much smaller size, equal to just the multiplicities $M_{RQ}$ of $Q$ in
$R^{\otimes m}$ \cite{MMMkn1,MMMkn2}:
\be
H_R^{\cal K}(G) = {\mathfrak N}^{w({\cal K})}\sum_{Q\in R^{\otimes m}} C_{RQ}^{\cal K} \chi_Q(G), \nn \\
\boxed{
C_{RQ}^{(a_{11}\ldots a_{1,m-1}|\ldots)} = \tr_{_{\!M_{RQ}}}\ \hat{\cal R}_{(1)}^{a_{11}}\ldots
\hat{\cal R}_{(m-1)}^{a_{1,m-1}}\ \hat{\cal R}_{(1)}^{a_{21}}\ldots
\hat{\cal R}_{(m-1)}^{a_{2,m-1}} \ldots
}
\label{RQdeco}
\ee
Here ${\mathfrak N}$ is a normalization factor which emerges due to our choice of non-standard normalization
of ${\cal R}$-matrices and $w({\cal K})$ is the writhe number.
The standard normalization of ${\cal R}$-matrix in the vertical framing
is restored with the factor $q^{-2\varkappa_R}$, where $\varkappa_R=\sum_{i,j\in R}(j-i)$ and the sum runs
over the Young diagram corresponding to $R$. In order to restore the topological invariance, one has to change framing
with a factor of $A^{-|R|}q^{-2\varkappa_R}$ which totally gives ${\mathfrak N}=A^{-|R|}q^{-4\varkappa_R}$.

What is important, in formula (\ref{RQdeco}) the knot and group dependencies are separated
and one can consider $H_R$ as a function of $A=q^N$ rather than $N$:
the parameter $N$ enters only the quantum dimensions $\chi_Q$
and can be easily substituted by $A$.
Moreover, this formula actually introduces the {\it extended} HOMFLY polynomial $H_R\{p\}$,
if $\chi_Q$ are interpreted as characters, which are  functions of
infinitely many time-variables $\{p_k\}$ instead of $N$ or $A$ \cite{MMMkn1,MMMkn2}.
The topological invariance is, however, lost
beyond the topological locus  $p_k=p_k^*=\frac{A^k-A^{-k}}{q^k-q^{-k}}$.

To deal with this formula one needs an explicit expression for
the $\hat{\cal R}$-matrices, see \cite{MMMkn2,IMMM13}.
Those papers contain many various observations about the structure
of $\hat{\cal R}$-matrices, and they were used to calculate many
non-trivial knot polynomials, still a complete description is not
yet found on that way, except for the fundamental
representation case of $R=[1]=\Box$, fully described in \cite{AnoMMM}, and for the (anti)symmetric
representation case \cite{IMMMfe,IMMM13}
(some results in this latter case are also reproduced by alternative methods \cite{iGFS}).

It is therefore natural to attack the case of arbitrary $R$ with the help
of the cabling approach \cite{cabling} and apply the results of \cite{AnoMMM}.
This is successfully done in \cite{fe21,AnoM}, this letter being a short
summary of \cite{AnoMMM} and \cite{AnoM},
purified from all the details
and extensive list of examples evaluated there.

Note also that here we restrict our discussion to the knots only. The links can be dealt with similarly, however,
they require some technical complications, thus, for the sake of brevity, we skip this extension (see details in \cite{AnoM}).

\section{$\hat{\cal R}$-matrices via paths in the representation graph \label{sop}}

Since the cabling reduces the problem from $R$ to the case of the fundamental representation,
the results of \cite{AnoMMM} are directly applicable \cite{AnoM}
and we begin from repeating them in a concise and pictorial form.

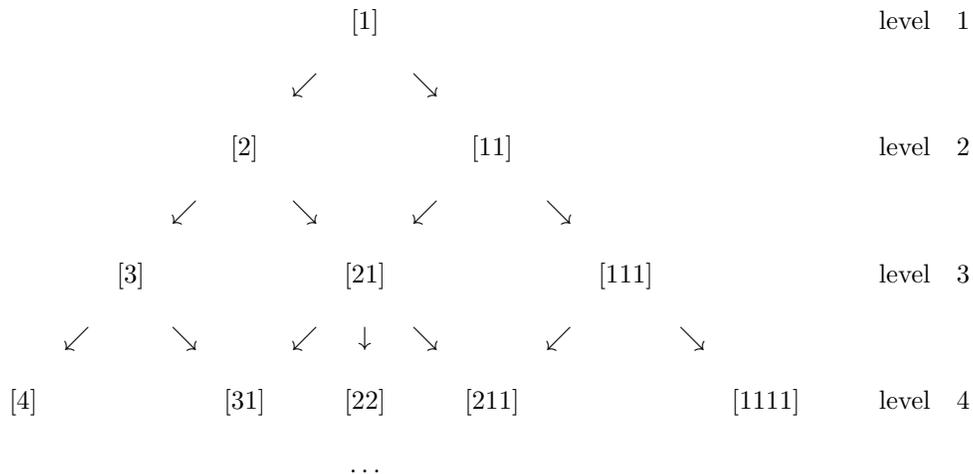
\be
\begin{array}{cccccccccccccccccc}
& &&& &&& &&& &&&    \\
& &&& &&& [1] &&& &&&    &&& {\rm level} & 1 \\
& &&& &&& &&& &&&    \\
& &&& &&\swarrow&&\searrow& &&& &    \\
& &&& &&& &&& &&&    \\
& &&& &[2]& &&&[11] &&&&   &&& {\rm level} & 2 \\
& &&& &&& &&& &&&    \\
& &&& \swarrow&&\searrow && \swarrow&&\searrow &&& &     \\
& &&& &&& &&& &&&    \\
& &&[3]&& && [21] && &&[111]&&   &&& {\rm level} & 3  \\
& &&& &&& &&& &&&    \\
& &\swarrow&&\searrow& &\swarrow&\downarrow&\searrow& &\swarrow&&\searrow&  \\
& &&& &&& &&& &&&    \\
& [4] &&&& [31] && [22]&& [211]&&&&[1111]   &&& {\rm level} & 4   \\
& &&& &&& &&& &&&    \\
& &&& &&&\ldots &&& &&&    \\
\end{array}
\nn
\ee

\bigskip

\bigskip

Implicit in \cite{AnoMMM} is representation of the coefficients $C_{\Box Q}$
in the form of a sum over paths in the representation graph \cite{AnoM}.
The first four levels of the representation graph of \cite{MMMkn2} are shown in the Figure:
in an obvious way it describes  the multiplication of fundamental representations [1].
The multiplicity $M_{\Box Q}$ of the representation $Q$ in $\Box^{\otimes |Q|}$
is obviously equal to the number of directed paths in the representation graph,
connecting $\Box$ and $Q$.
More generally, $M_{RQ}$ is equal to the number of directed paths between $R$ and $Q$.
The matrices $\hat {\cal R}_{(i)}$ , $i=1,\ldots,m -1$
can be represented in the basis
of paths between $Q$ and $\Box$ and according to \cite{AnoMMM,AnoM}
they have extremely simple form in this basis.

First of all, with each index $i$ of the matrix $\hat {\cal R}_{(i)}$ one associates a level $i$ in the graph.
A given path ${\cal P}$ is passing through exactly one vertex $P_i$ at level $i$
and through some two adjacent vertices $P_{i-1}$ and $P_{i+1}$ at levels
$i-1$ and $i+1$.
The structure of the representation graph is such that these $P_{i-1}$ and $P_{i+1}$
are connected either by a single two-segment path (singlet)
(then it is a fragment of our $P$) or by two such paths (doublet), the segments of
our path ${\cal P}$ and another path ${\cal P}'$.
We call the transformations
${\cal P} \leftrightarrow {\cal P}'$ a {\it flip}.

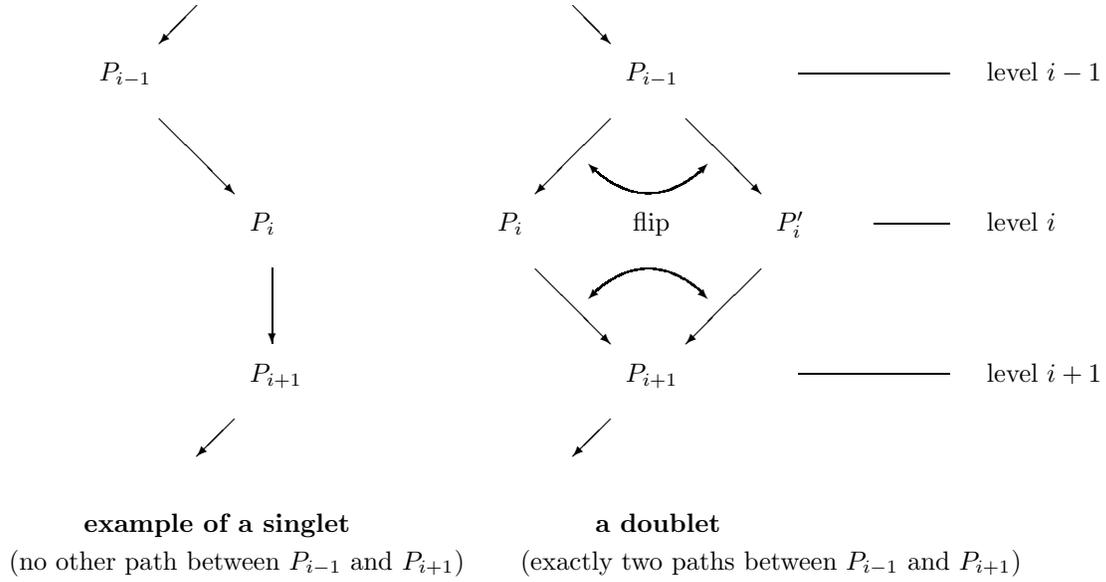
\begin{picture}(120,88)(-90,-73)
\put(-60,10){\vector(-1,-1){5}}
%
%\put(-75,-5){\vector(-1,-1){10}}
\put(-65,-5){\vector(1,-1){10}}
\put(-50,-25){\vector(0,-1){10}}
%\put(15,-25){\vector(-1,-1){10}}
%
\put(-73,0){\mbox{$P_{i-1}$}}
%\put(-20,-20){\mbox{$P_{i}$}}
\put(-53,-20){\mbox{$P_{i}$}}
\put(-53,-40){\mbox{$P_{i+1}$}}
\put(-55,-45){\vector(-1,-1){5}}
\put(-75,-60){\mbox{{\bf example of a singlet}}}
\put(-85,-65){\mbox{(no other path between $P_{i-1}$ and $P_{i+1}$)}}
\put(-7,-60){\mbox{{\bf a doublet}}}
\put(-17,-65){\mbox{(exactly two paths between $P_{i-1}$ and $P_{i+1}$)}}
\put(-10,10){\vector(1,-1){5}}
\put(-5,-5){\vector(-1,-1){10}}
\put(5,-5){\vector(1,-1){10}}
\put(-15,-25){\vector(1,-1){10}}
\put(15,-25){\vector(-1,-1){10}}
\put(-3,0){\mbox{$P_{i-1}$}}
\put(-20,-20){\mbox{$P_{i}$}}
\put(17,-20){\mbox{$P'_{i}$}}
\put(-3,-40){\mbox{$P_{i+1}$}}
\put(-5,-45){\vector(-1,-1){5}}
\qbezier(-7,-12)(0,-18)(7,-12)
\put(-2,-20){\mbox{flip}}
\qbezier(-7,-28)(0,-22)(7,-28)
\put(-6,-13){\vector(-1,1){2}}
\put(6,-13){\vector(1,1){2}}
\put(-6,-27){\vector(-1,-1){2}}
\put(6,-27){\vector(1,-1){2}}
\put(45,0){\mbox{level $i-1$}}
\put(45,-20){\mbox{level $i$}}
\put(45,-40){\mbox{level $i+1$}}
\put(40,1){\line(-1,0){20}}
\put(40,-19){\line(-1,0){10}}
\put(40,-39){\line(-1,0){20}}
\end{picture}

In the former case (singlet) our path ${\cal P}$ provides
a diagonal element in $\hat {\cal R}_{(i)}$
and it is equal to either $q$ or $-1/q$.
In the language of Young diagrams the singlet appears when the two boxes
added to the diagram $P_{i-1}$ in order to form $P_{i+1}$ lie either
in the same row, then we put $q$ at the diagonal of $\hat{\cal R}_i$;
or in the same column, then we put $-1/q$.

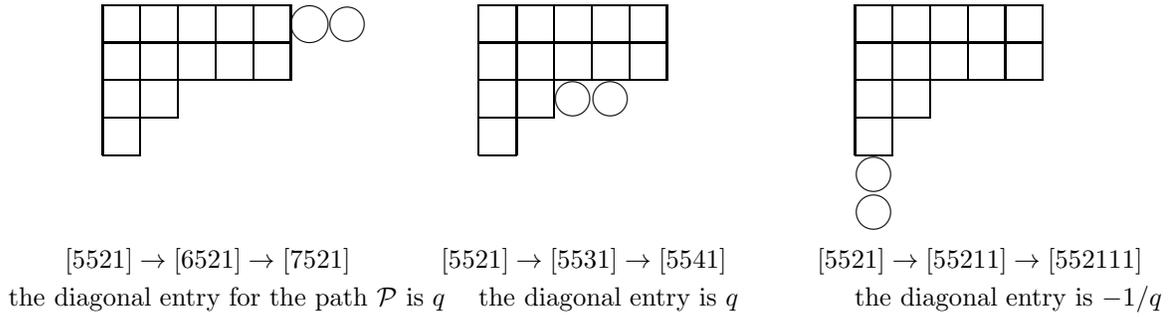
\begin{picture}(100,60)(-20,-53)
\unitlength 0.5mm % = 2.845pt
\put(0,0){\line(0,-1){40}}
\put(10,0){\line(0,-1){40}}
\put(20,0){\line(0,-1){30}}
\put(30,0){\line(0,-1){20}}
\put(40,0){\line(0,-1){20}}
\put(50,0){\line(0,-1){20}}
\put(0,0){\line(1,0){50}}
\put(0,-10){\line(1,0){50}}
\put(0,-20){\line(1,0){50}}
\put(0,-30){\line(1,0){20}}
\put(0,-40){\line(1,0){10}}
\put(65,-5){\circle{9}}
\put(55,-5){\circle{10}}
\put(-10,-70){\mbox{$[5521]\rightarrow [6521] \rightarrow [7521]$}}
\put(-25,-80){\mbox{the diagonal entry for the path ${\cal P}$ is $q$}}
\put(95,-95){\mbox{{\bf examples of singlets}}}
\put(100,0){\line(0,-1){40}}
\put(110,0){\line(0,-1){40}}
\put(120,0){\line(0,-1){30}}
\put(130,0){\line(0,-1){20}}
\put(140,0){\line(0,-1){20}}
\put(150,0){\line(0,-1){20}}
\put(100,0){\line(1,0){50}}
\put(100,-10){\line(1,0){50}}
\put(100,-20){\line(1,0){50}}
\put(100,-30){\line(1,0){20}}
\put(100,-40){\line(1,0){10}}
\put(125,-25){\circle{9}}
\put(135,-25){\circle{9}}
\put(90,-70){\mbox{$[5521]\rightarrow [5531] \rightarrow [5541]$}}
\put(100,-80){\mbox{the diagonal entry is $q$}}
\put(200,0){\line(0,-1){40}}
\put(210,0){\line(0,-1){40}}
\put(220,0){\line(0,-1){30}}
\put(230,0){\line(0,-1){20}}
\put(240,0){\line(0,-1){20}}
\put(250,0){\line(0,-1){20}}
\put(200,0){\line(1,0){50}}
\put(200,-10){\line(1,0){50}}
\put(200,-20){\line(1,0){50}}
\put(200,-30){\line(1,0){20}}
\put(200,-40){\line(1,0){10}}
\put(205,-45){\circle{9}}
\put(205,-55){\circle{9}}
\put(190,-70){\mbox{$[5521]\rightarrow [55211] \rightarrow [552111]$}}
\put(200,-80){\mbox{the diagonal entry is $-1/q$}}
\end{picture}

In the latter case (the doublet) the two boxes are neither in the same row nor in the same
column, and the two paths ${\cal P}$ and ${\cal P}'$ form a $2\times 2$ block in
$\hat {\cal R}_{(i)}$. This block is described as follows.
First, that of the paths ${\cal P}$ and ${\cal P}'$ which lies to the left of the other,
corresponds to the left column and to the first row of the
$2\times 2$ block.  Second, the Young diagrams $P_{i+1}$ is obtained
by adding two boxes to the diagram $P_{i-1}$, and the two paths correspond
to doing this in two different orders, thus providing
at the intermediate stage the two adjacent vertices $P_{i}$ and $P'_{i}$.
The two added boxes are connected
by a hook in the Young diagram, which has length $n$
(measured between the {\it centers} of the two boxes).
Then the $2\times 2$ block is equal to
\be
\left(\begin{array}{ccc}
-q^{-n} c_n &\ & s_n \\ \\
s_n &  & q^n c_n
\end{array}
\right),
\ \ \ \ \ \ c_n = \frac{1}{[n]_q} = \frac{q-q^{-1}}{q^n-q^{-n}}, \ \ \ \ \ \
s_n = \sqrt{1-c_n^2}= \frac{\sqrt{[n-1]_q[n+1]_q}}{[n]_q}
\ee
The following picture shows that for
$P_{i-1} = [5521]$, $P_{i+1}= [6522]$, $P_i = [6521]$ and $P'_i = [5522]$
the parameter $n= 7$:

\begin{picture}(100,70)(-50,-65)
\unitlength 0.7mm % = 2.845pt
\put(0,0){\line(0,-1){40}}
\put(10,0){\line(0,-1){40}}
\put(20,0){\line(0,-1){30}}
\put(30,0){\line(0,-1){20}}
\put(40,0){\line(0,-1){20}}
\put(50,0){\line(0,-1){20}}
\put(0,0){\line(1,0){50}}
\put(0,-10){\line(1,0){50}}
\put(0,-20){\line(1,0){50}}
\put(0,-30){\line(1,0){20}}
\put(0,-40){\line(1,0){10}}
\put(15,-35){\circle{9}}
\put(55,-5){\circle*{10}}
\put(15,-35){\line(0,1){30}}
\put(55,-5){\line(-1,0){40}}
\put(-6,-60){\mbox{$[5521] $}}
\put(22,-70){\mbox{$[6521]$}}
\put(22,-50){\mbox{$[5522]$}}
\put(50,-60){\mbox{$[6522]$}}
\put(12,-56){\vector(2,1){6}}
\put(12,-64){\vector(2,-1){6}}
\put(40,-52){\vector(2,-1){6}}
\put(40,-68){\vector(2,1){6}}
\put(45,-34){\mbox{$n=7$}}
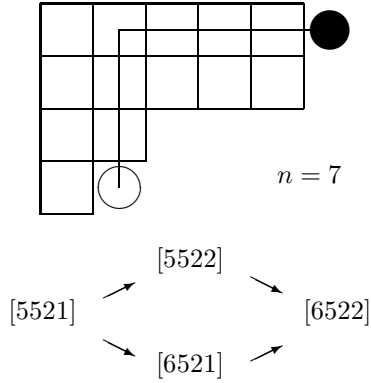
\put(0,-85){\mbox{{\bf example of the doublet}}}
\end{picture}

This provides a complete and very explicit description of the matrices $\hat{\cal R}_{(i)}$,
$i=1,\ldots,m -1$, which appear under the trace in the expression for
${C}_{\Box Q}$.
Note that in this form these matrices depend on $Q$.
This description can seem rather lengthy and tricky,
but in practice it is very algorithmic, easily programmable
and very effective for practical calculations \cite{AnoM}.

\bigskip

The paths from $\Box$ to $Q$ in the representation graph are labeled
by the standard Young tableaux, i.e. the Young diagram $Q$
with the numbers $1,\ldots,|Q|$ assigned to its boxes in such a way
that the emerging sequences are increasing along each row and each
column. The number in a given box is just the number of steps, at which
the box was attached to the diagram when moving from $\Box$ to $Q$.
The number of paths is therefore equal to the well-known number
of the standard Young tableaux,
\be
M_{\Box Q} = \frac{|Q|!}{\prod_{{\rm boxes}\in Q} L({\rm box})}
\ee
where $L({\rm box})$ is the length of the hook in $Q$ associated
to the box.

{\bf
Thus the central formula (\ref{RQdeco}) can be considered as a multiple
sum over the standard Young tableaux of $Q$, and the number of summations
is equal to the number $\sum_{ij} |a_{ij}|$ of vertices in the braid.
}

\section{Examples \label{Exa}}

To illustrate the use of these formulas we calculate the contributions
of simple diagrams $Q$ to the simple HOMFLY polynomials.
There is nothing new in these formulas, they are present here just to illustrate how
the method works. For numerous new examples see \cite{AnoM}.

If one considers only $H_\Box$ in the fundamental representation $R=\Box=[1]$, $|\Box|=1$,
then (\ref{Ccab}) for an $m$-strand braid involves $Q$ of the size $m$,
i.e. $Q$ at the level $m$ in the representation graph.
To make use of (\ref{Ccab}) one needs to know the $m-1$ matrices $\hat{\cal R}_{(i)}$.

\bigskip

\subsection*{m=2:}

There are two possible $Q=[2]$ and $Q=[11]$ and one matrix $\hat{\cal R}_{(1)}$
in each case.

\bigskip

For $\underline{Q=[2]}$ there is a single path from $\Box=[1]$ to $[2]$, i.e. $M_{\Box [2]}=1$,
it coincides
with the segment $([1],[2])$, thus, according to our rules the
corresponding $1\times 1$ matrix $\hat{\cal R}_{(1)}=q$.

\bigskip

Similarly for $\underline{Q=[11]}$ one gets $\hat{\cal R}_{(1)}=-1/q$.

\bigskip

Thus the generic expression (\ref{HRcab}) in this case is
\be
A^nH_\Box^{(n)} = q^n \chi_{[2]}(G) + (-q)^{-n}\chi_{[11]}(G), \ \ \ \ \ \ \  n\ {\rm odd}
\label{2stra}
\ee
The coefficient at the l.h.s. takes into account the normalization factor ${\mathfrak N}$.

\subsection*{m=3:}

For \underline{$Q=[3]$ and $[111]$} there are unique paths from $\Box$
and the corresponding matrices $\hat{\cal R}_{(1)}=\hat{\cal R}_{(2)}$
are again $1\times 1$ and are equal to $q$ and $-1/q$ respectively.

\bigskip

However, for $\underline{Q=[21]}$ the situation is already different.
There are two paths between $\Box$ and $[21]$, and the left one,
$[1]\rightarrow [2]\rightarrow [21]$ contains the segment $([1],[2])$, while the second
path $[1]\rightarrow [11]\rightarrow [21]$ contains the segment $([1],[11])$.
This means that the matrix $\hat{\cal R}_{(1)}$ in the sector $Q=[21]$
is $2\times 2$, it is diagonal with the entries $q$ and $-1/q$.
The matrix $\hat{\cal R}_{(2)}$ is again $2\times 2$, but it is not diagonal,
because the two paths are connected exactly by the {\it flip}.
Since the length of the hook in this case is $n=2$,
our rules imply that $\hat{\cal R}_{(2)} =
\frac{1}{[2]_q}\left(\begin{array}{cc} -1/q^{2} & \sqrt{[3]_q} \\
\sqrt{[3]_q} &  q^2 \end{array}\right)$
and the formula for the arbitrary $3$-strand braid is \cite{MMMkn2}

$$
A^{a_1+b_1+a_2+b_2+\ldots} H_\Box^{(a_1b_1|a_2b_2|\ldots}
= q^{a_1+b_1+a_2+b_2+\ldots}\ \chi_{[3]}(G)\ +\
(-1/q)^{a_1+b_1+a_2+b_2+\ldots}\ \chi_{[111]}(G)\ +
$$
\be
+ \tr_{2\times 2}\left\{
\left(\begin{array}{cc} q & 0 \\ \\ 0 & -1/q \end{array}\right)^{a_1}
\left(\begin{array}{cc} -\frac{1}{q^2[2]_q} & \frac{\sqrt{[3]_q}}{[2]_q} \\ \\
\frac{\sqrt{[3]_q}}{[2]_q} &  \frac{q^2}{[2]_q} \end{array}\right)^{b_1}
\left(\begin{array}{cc} q & 0 \\ \\ 0 & -1/q \end{array}\right)^{a_2}
\left(\begin{array}{cc} -\frac{1}{q^2[2]_q} & \frac{\sqrt{[3]_q}}{[2]_q} \\ \\
\frac{\sqrt{[3]_q}}{[2]_q} &  \frac{q^2}{[2]_q} \end{array}\right)^{b_2}
\ldots \right\}
\label{3stra}
\ee

\begin{picture}(100,50)(-50,-40)
\put(0,0){\line(1,0){10}}
\put(0,-10){\line(1,0){10}}
\put(0,0){\line(0,-1){10}}
\put(10,0){\line(0,-1){10}}
\put(15,-5){\circle*{9}}
\put(5,-15){\circle{9}}
\put(5,-5){\line(0,-1){10}}
\put(5,-5){\line(1,0){10}}
\put(0,-30){\mbox{$[1]\rightarrow [21]$}}
\put(3,-37){\mbox{$n=2$}}
\put(50,0){\line(1,0){20}}
\put(50,-10){\line(1,0){20}}
\put(50,0){\line(0,-1){10}}
\put(60,0){\line(0,-1){10}}
\put(70,0){\line(0,-1){10}}
\put(75,-5){\circle*{9}}
\put(55,-15){\circle{9}}
\put(55,-5){\line(0,-1){10}}
\put(55,-5){\line(1,0){20}}
\put(50,-30){\mbox{$[2]\rightarrow [31]$}}
\put(53,-37){\mbox{$n=3$}}
\end{picture}

\subsection*{m=4:}

For \underline{$Q=[4]$ and $[1111]$} all the three matrices $\hat{\cal R}_{1,2,3}$
are $1\times 1$ and equal to $q$ and $-1/q$ respectively.

\bigskip

For $\underline{Q=[31]}$ there are three paths. Ordered from the left to the right they are:
$$
\alpha = [1]\rightarrow [2]\rightarrow [3]\rightarrow [31], \ \ \ \
\beta = [1]\rightarrow [2]\rightarrow [21]\rightarrow [31], \ \ \ \
\gamma = [1]\rightarrow [11]\rightarrow [21]\rightarrow [31]
$$
At level $2$ the flip relates $\beta$ and $\gamma$,
at level $3$ relates $\alpha$ and $\beta$.
This implies that in sector $[31]$ one has, \cite{MMMkn2}
\be
\hat{\cal R}_{(1)} = \left(\begin{array}{ccc}
q &&\\ & q & \\ && -1/q \end{array}\right), \ \ \
\hat{\cal R}_{(2)} = \left(\begin{array}{ccc}
q &0&0\\ \\ 0 & -\frac{1}{q^2[2]_q} & \frac{\sqrt{[3]_q}}{[2]_q} \\ \\
0& \frac{\sqrt{[3]_q}}{[2]_q} & \frac{q^2}{[2]_q} \end{array}\right), \ \ \
\hat{\cal R}_{(3)} = \left(\begin{array}{ccc}
-\frac{1}{q^3[3]_q} &\frac{\sqrt{[2]_q[4]_q}}{[3]_q}&0\\ \\
\frac{\sqrt{[2]_q[4]_q}}{[3]_q} &\frac{q^3}{[3]_q} & 0\\ \\ 0&0& q \end{array}\right)
\label{R31}
\ee

\bigskip

For $\underline{Q=[211]}$ the answer is similar:
the three paths are now
$$
\bar\alpha = [1]\rightarrow [2]\rightarrow [21]\rightarrow [211], \ \ \ \
\bar\beta = [1]\rightarrow [11]\rightarrow [21]\rightarrow [211], \ \ \ \
\bar\gamma= [1]\rightarrow [11]\rightarrow [111]\rightarrow [211]
$$
and
\be
\hat{\cal R}_{(1)} = \left(\begin{array}{ccc}
q &&\\ & \!\!\!-1/q & \\ && \!\!\!\! -1/q \end{array}\right)\!, \ \
\hat{\cal R}_{(2)} = \left(\begin{array}{ccc}
-\frac{1}{q^2[2]_q} &\frac{\sqrt{[3]_q}}{[2]_q} &0\\ \\
\frac{\sqrt{[3]_q}}{[2]_q} & \frac{q^2}{[2]_q} &0  \\ \\
0& 0 &\!\!\!\! -1/q \end{array}\right)\!, \ \
\hat{\cal R}_{(3)} = \left(\begin{array}{ccc}
-1/q\!\!\! &0&0 \\ \\ 0& -\frac{1}{q^3[3]_q} & \frac{\sqrt{[2]_q[4]_q}}{[3]_q}\\ \\
0& \frac{\sqrt{[2]_q[4]_q}}{[3]_q} & \frac{q^3}{[3]_q}   \end{array}\right)
\label{R211}
\ee

\bigskip

For $\underline{Q=[22]}$ there are just two paths,
$$
\delta = [1]\rightarrow [2]\rightarrow [21] \rightarrow [22], \ \ \ \
\bar\delta = [1]\rightarrow [11]\rightarrow [21] \rightarrow [22]
$$
and they are related by the flip at level $2$.
This means that both $\hat{\cal R}_{(1)}$ and $\hat{\cal R}_{(3)}$ are
$2\times 2$, and diagonal, while $\hat{\cal R}_{(2)}$ is not:
\be
\hat{\cal R}_{(1)} = \hat{\cal R}_{(3)} =
\left(\begin{array}{cc} q & 0 \\ \\ 0 & -1/q \end{array}\right), \ \ \ \
\hat{\cal R}_{(2)} =
\left(\begin{array}{cc} -\frac{1}{q^2[2]_q} & \frac{\sqrt{[3]_q}}{[2]_q} \\ \\
\frac{\sqrt{[3]_q}}{[2]_q} &  \frac{q^2}{[2]_q} \end{array}\right)
\label{R22}
\ee

The counterpart of (\ref{2stra}) and (\ref{3stra}) is now obvious, but rather lengthy,
so we do not write it down here.

\section{Cabling method}

Cabling is based on the fact that the representation $R$ appears in the product
of $|R|$ fundamental representations, $R\in \Box^{|R|}$, thus, the answer for $H_R^{\cal K}$
can be extracted from the answer for $H_{\Box^{\otimes |R|}}^{{\cal K}^{|R|}}$
by the projection:
\be
H_R^{\cal K}(G) ={\mathfrak N}_c
\sum_{Q\in \Box^{\otimes m|R|}} {\mathfrak  C}_{\Box Q}^{{\cal K}^{|R|}} \chi_Q(G)
\label{HRcab}
\ee
where the normalization factor is slightly corrected as compared with (\ref{RQdeco}), see \cite{AnoM} for details.
Cabling is widely used not only in the knot theory \cite{cabling},
but also in the theory of ${\cal R}$-matrices and integrable systems
\cite{cabint} (where it is called the fusion procedure).

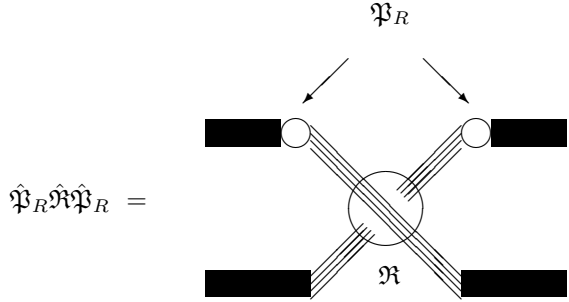
\begin{picture}(100,50)(-80,-20)
\linethickness{10.0pt}
\put(-24,10){\line(1,0){10}}
\put(-24,-10){\line(1,0){14}}
\put(14,10){\line(1,0){10}}
\put(10,-10){\line(1,0){14}}
\put(-12,10){\circle{4}}
%\put(-12,-10){\circle{4}}
\put(12,10){\circle{4}}
%\put(-12,-10){\circle{4}}
%
\put(-10,11){\line(1,-1){20}}
\put(-10,10){\line(1,-1){20}}
\put(-10,9){\line(1,-1){20}}
\put(-10,8){\line(1,-1){20}}
\put(-10,-9){\line(1,1){7}}
\put(-10,-10){\line(1,1){7.5}}
\put(-10,-11){\line(1,1){8}}
\put(-10,-12){\line(1,1){8.5}}
\put(10,11){\line(-1,-1){8.5}}
\put(10,10){\line(-1,-1){8}}
\put(10,9){\line(-1,-1){7.5}}
\put(10,8){\line(-1,-1){7}}
\linethickness{0.4pt}
\put(-2,25){\mbox{${\mathfrak P}_R$}}
\put(-5,20){\vector(-1,-1){6}}
\put(5,20){\vector(1,-1){6}}
\put(0,0){\circle{10}}
\put(-1,-10){\mbox{${\mathfrak R}$}}
\put(-50,0){\mbox{$\hat{\mathfrak P}_R\hat{\mathfrak R}\hat{\mathfrak P}_R\ =$}}
\end{picture}

\noindent
Here ${\cal K}^{|R|}$ is an $m|R|$-strand braid, obtained from ${\cal K}$
by substituting each line (strand) by a bunch (cable) of $|R|$ strands
so that the intersection of two strands is now a peculiar combination ${\mathfrak R}$ of $|R|^2$
original ${\cal R}$-matrices, but in the fundamental representations.
In other words,
\be
\boxed{
{\mathfrak C}_{\Box Q}^{{\cal K}^{|R|}} = \tr_{_{\!M_{\Box Q}}}\
\hat{\mathfrak P}_R\  \hat{\mathfrak R}_{(1)}^{a_{11}}\ldots
\hat{\mathfrak  R}_{(m-1)}^{a_{1,m-1}}\ \hat{\mathfrak R}_{(1)}^{a_{21}}\ldots
\hat{\mathfrak  R}_{(m-1)}^{a_{2,m-1}} \ldots
}
\label{Ccab}
\ee
${\mathfrak P}_R$ is the projection from the reducible representation $\Box^{\otimes |R|}$ onto $R$.
It can actually be inserted everywhere in between the ${\mathfrak R}$
matrices, but for the case of {\it knots} (not links) a single insertion is sufficient.

Now, (\ref{Ccab}) can be written explicitly and calculated
with the help of  sec.\ref{sop},
but in terms of the $m|R|$-strand braid and
the corresponding $\hat {\cal R}$ matrices replaced with
$\hat{\mathfrak R}$.
This is a rather cumbersome expression even for
the simplest knots (therefore we do not rewrite (\ref{Ccab}) in this way)
but absolutely straightforward and adequate for practical computations.

Moreover, here comes a bonus of the path sum representation:
{\bf the projector ${\mathfrak P}_R$ is nearly trivial}, at least in the case of knots:
one should include only the directed paths from $Q$ to $\Box$,
which pass through the vertex $R$, this effectively decreases the size
of the matrices $\hat {\mathfrak R}_{(I)}$, $I=1,\ldots,m-1$
from $M_{\Box Q}$ to $M_{\Box R}\cdot M_{RQ}$.
{\bf NB:} The constituent matrices $\hat{\cal R}_{(i)}$, $i=1,\ldots,m|R|-1$
can {\it not} be reduced in this way: one can {\it not} insert the
projector {\it inside}  $\hat {\mathfrak R}$.

\section{Cabling for m=2 and $|$R$|$=2:}

Now we can use the last example in s.\ref{Exa}
to demonstrate how the cabling works in our formulas.
The $4$-strand braids are enough to describe only the $2$-strand (torus)
knots in representations $[2]$ and $[11]$.
In this case there is just a single combined matrix
\be
\hat{\mathfrak R}_{(1)} =
\hat{\cal R}_{(2)}\hat{\cal R}_{(1)}\hat{\cal R}_{(3)}\hat{\cal R}_{(2)}
\ee

According to our rules,
if we consider, for example, $H_{[2]}^{(n)}$ for a $2$-strand knot,
we should leave in $\hat{\mathfrak R}_{(1)}$ only the paths passing through the vertex $[2]$.

\bigskip

This means that the single path leading to $\underline{Q=[4]}$ remains intact,
while the one to $[1111]$ is now eliminated; this means that there
will be no contribution of $\underline{Q=[1111]}$ to $H_{[2]}^{(n)}$.

\bigskip

From the three paths which led to $\underline{Q=[31]}$ only two remain, $\alpha$ and $\beta$,
thus the third line and the third column  should be omitted from the matrix $\hat{\mathfrak R}$:
$$
\hat {\mathfrak R} =
\left(\begin{array}{ccc}
q &0&0\\ \\ 0& -\frac{1}{q^2[2]_q} & \frac{\sqrt{[3]_q}}{[2]_q} \\ \\
0& \frac{\sqrt{[3]_q}}{[2]_q} & \frac{q^2}{[2]_q} \end{array}\right)
\left(\begin{array}{ccc}
q &&\\ & q & \\ && -1/q \end{array}\right)
\left(\begin{array}{ccc}
-\frac{1}{q^3[3]_q} &\frac{\sqrt{[2]_q[4]_q}}{[3]_q}&0\\ \\
\frac{\sqrt{[2]_q[4]_q}}{[3]_q} &\frac{q^3}{[3]_q} &0 \\ \\ 0&0& q \end{array}\right)
\left(\begin{array}{ccc}
q &0&0\\ \\ 0& -\frac{1}{q^2[2]_q} & \frac{\sqrt{[3]_q}}{[2]_q} \\ \\
0 & \frac{\sqrt{[3]_q}}{[2]_q} & \frac{q^2}{[2]_q} \end{array}\right)
=
$$
$$
= \left(\begin{array}{ccc}-\frac{1}{[3]_q}&-\frac{\sqrt{[2]_q[4]_q}}{[2]_q[3]_q}&
\frac{q^2\sqrt{[2]_q[3]_q[4]_q}}{q^2[2]_q[3]_q}\\ \\
-\frac{\sqrt{[2]_q[4]_q}}{[2]_q[3]_q}&-\frac{[4]_q}{[2]_q[3]_q}&-\frac{q^2}{\sqrt{[3]_q}}\\ \\
\frac{q^2\sqrt{[2]_q[3]_q[4]_q}}{[2]_q[3]_q}&-\frac{q^2}{\sqrt{[3]_q}}&0
\end{array}\right)\ \longrightarrow\
$$
\be
\longrightarrow\ \hat{\mathfrak P}_{[2]}\hat {\mathfrak R}\hat{\mathfrak P}_{[2]}  =
\left(\begin{array}{ccc} 1 &&\\&1&   \\&&0   \end{array}\right)
\hat {\mathfrak R}
\left(\begin{array}{ccc}1\\&1\\&&0\end{array}\right)\ \longrightarrow\
\left(\begin{array}{cc}-\frac{1}{[3]_q}&-\frac{\sqrt{[2]_q[4]_q}}{[2]_q[3]_q}\\ \\
-\frac{\sqrt{[2]_q[4]_q}}{[2]_q[3]_q}&-\frac{[4]_q}{[2]_q[3]_q}
\end{array}\right)
\ee
This $2\times 2$ matrix has two eigenvalues: $0$ and $-1$.

\bigskip

Similarly from the three paths to $\underline{Q=[211]}$ only one survives, $\bar\alpha$,
and $\hat{\mathfrak R}$ matrix is reduced to $1\times 1$:

$$
\hat {\mathfrak R} =
\left(\begin{array}{ccc}
-\frac{1}{q^2[2]_q} &\frac{\sqrt{[3]_q}}{[2]_q} &0\\ \\
\frac{\sqrt{[3]_q}}{[2]_q} & \frac{q^2}{[2]_q} & 0 \\ \\
0& 0 & \!\!\! -1/q \end{array}\right)
\left(\begin{array}{ccc}
q &&\\ &\!\!\!\! -1/q & \\ && \!\!\!\!-1/q \end{array}\right)
\left(\begin{array}{ccc}
-1/q\!\!\! &0& 0\\ \\ 0 & -\frac{1}{q^3[3]_q} & \frac{\sqrt{[2]_q[4]_q}}{[3]_q}\\ \\
0 & \frac{\sqrt{[2]_q[4]_q}}{[3]_q} & \frac{q^3}{[3]_q}   \end{array}\right)
\left(\begin{array}{ccc}
-\frac{1}{q^2[2]_q} &\frac{\sqrt{[3]_q}}{[2]_q} & 0 \\ \\
\frac{\sqrt{[3]_q}}{[2]_q} & \frac{q^2}{[2]_q} & 0 \\ \\
0 & 0 & -1/q \end{array}\right)=
$$
$$
= \left(\begin{array}{ccc}0&\frac{1}{q^2\sqrt{[3]_q}}&\frac{\sqrt{[2]_q[3]_q[4]_q}}{q^2[2]_q[3]_q}
\\ \\ \frac{1}{q^2\sqrt{[3]_q}}&-\frac{[4]_q}{[2]_q[3]_q}&\frac{\sqrt{[2]_q[4]_q}}{[2]_q[3]_q}\\ \\
\frac{\sqrt{[2]_q[3]_q[4]_q}}{q^2[2]_q[3]_q}&\frac{\sqrt{[2]_q[4]_q}}{[2]_q[3]_q}&
-\frac{1}{[3]_q}\end{array}\right)\ \longrightarrow\
$$
\be
\longrightarrow\ \hat{\mathfrak P}_{[2]}\hat {\mathfrak R}\hat{\mathfrak P}_{[2]} =
\left(\begin{array}{ccc} 1 &&\\&0& \\ &&0   \end{array}\right)
\hat {\mathfrak R}
\left(\begin{array}{ccc}1\\&0\\&&0\end{array}\right)\ \longrightarrow\ \Big(0\Big)
\ee
Thus, despite a path survives in the $[211]$ sector, its contribution
is actually vanishing, as it should be, because $[211]\notin [2]\otimes[2]$.

\bigskip

Also for $\underline{Q=[22]}$ only one path of the two remains, $\delta$, and
$$
\left(\begin{array}{cc} -\frac{1}{q^2[2]_q} & \frac{\sqrt{[3]_q}}{[2]_q} \\ \\
\frac{\sqrt{[3]_q}}{[2]_q} &  \frac{q^2}{[2]_q} \end{array}\right)
\left(\begin{array}{cc} q & 0 \\ 0 & -1/q \end{array}\right)^2
\left(\begin{array}{cc} -\frac{1}{q^2[2]_q} & \frac{\sqrt{[3]_q}}{[2]_q} \\ \\
\frac{\sqrt{[3]_q}}{[2]_q} &  \frac{q^2}{[2]_q} \end{array}\right)
= \left(\begin{array}{cc}
\frac{1}{q^2} & 0 \\ \\ 0 & q^2
\end{array}\right)
$$
\be
\longrightarrow\
\hat{\mathfrak P}_{[2]}\hat {\mathfrak R}\hat{\mathfrak P}_{[2]} =
\left(\begin{array}{cc} 1 &\\&0     \end{array}\right)
\hat {\mathfrak R}
\left(\begin{array}{cc} 1 &\\&0     \end{array}\right) \ \longrightarrow \
\left( \frac{1}{q^2}\right)
\ee

Putting everything together one gets for odd $n$ (i.e. for the knot):
$$
q^{2n}A^{2n}H_{[2]}^{(n)} = \Tr^{\rm grad}_{[1]^{\otimes 4}} \hat{\mathfrak P}_{[2]}
\hat{\mathfrak R}^n =
\Tr^{\rm grad}_{[1]^{\otimes 4}} \Big(\hat{\mathfrak P}_{[2]}
\hat{\mathfrak R}\hat{\mathfrak P}_{[2]}\Big)^n =
$$

$$
= q^{4n} \chi_{[4]}(G)
+ \tr_{2\times 2} \left\{
\left(\begin{array}{cc}
-\frac{1}{[3]_q}&-\frac{\sqrt{[2]_q[4]_q}}{[2]_q[3]_q}\\ \\
-\frac{\sqrt{[2]_q[4]_q}}{[2]_q[3]_q}&-\frac{[4]_q}{[2]_q[3]_q}
\end{array}\right)
\right\}^n\chi_{[31]}(G)
+ q^{-2n}\chi_{[22]}(G) =
$$
\be
= q^{-2n}\Big(  q^{6n}\chi_{[4]}(G) - q^{2n} \chi_{[31]}(G) +  \chi_{[22]}(G)\Big)
\label{42stra}
\ee
The formula at the last line is the standard Rosso-Jones expression \cite{RJ}
for the colored HOMFLY polynomial in the case of a torus knot. \\
{\bf NB:} Note that the matrices $\hat{\mathfrak R}$ do {\it not}
commute with $\hat{\mathfrak P}_R$ and that the projector
is as simple as described above only for the first cable in the braid.\\
Note also that   two  of the five $Q\in [1]^4$ do not contribute to the middle line
for somewhat different reasons:
$\chi_{[1111]}$ does not appear, because there are no paths,
going from $\Box$ to $[1111]$ via $[2]$, while for $\chi_{[211]}$ such
path exists, just its contribution to the relevant element of $\hat{\mathfrak R}$
is zero.

\bigskip

These examples illustrate all the peculiarities of our formulas.
Once they are understood, the use of the path sum formalism
is straightforward.

\section{Open questions}

The path sum formula provides a complete solution
for the evaluation problem of arbitrary colored HOMFLY polynomials.
It is nice looking and theoretically attractive,
and it is algorithmic and very effective for practical computations
(a vast list of examples is provided in \cite{fe21,AnoM}).
It represents the knot polynomials in the form of a character expansion
\cite{MMMkn1,MMMkn2},
and therefore expresses them directly in terms of $N$, $A$, or
the time-variables $p_k$, whatever one prefers.
Still there are a few obvious directions to study.

First, our final formula for the colored knots
heavily relies on the cabling approach and
therefore is in a certain sense more involved than (\ref{RQdeco}).
In particular, when $R\neq \Box$,
the sizes of ${\cal R}$-matrices are considerably
bigger than theoretically possible
(though the entries are simpler).
Of course, {\bf the path sum representation exists for the arbitrary $R$},
but it is in terms of another representation graph, $\Gamma_R$ describing
powers of the representation $R$, and the corresponding ${\cal R}$-matrices
are more involved (their constituents are no longer just
$1\times 1$ and $2\times 2$ blocks).
At the same time that graph is the subset of the full one, $\Gamma_\Box$,
which we considered in the present letter, and
the sum over the paths from $Q$ to $\Box$,
passing through the vertex $R$,
can be also considered as a sum over just
the paths from $Q$ to $R$, more in the spirit of
eq.(\ref{RQdeco}).
However, there are many such paths in $\Gamma_\Box$
with the same image in $\Gamma_R$, and they contribute
to the matrix elements of the composite $\hat{\mathfrak R}$.
The quantities (\ref{Ccab}) contain also multiple sums
over the paths from $R$ to $\Box$, but in fact they do not depend
on this, and one can simply fix one such path arbitrarily
(as a kind of a gauge fixing).
It would be very interesting to find a relevant
modification of the path sum formula which would
not refer to the cabling procedure, and \cite{IMMM13} strongly
implies that this can be possible and that
such a formula should possess its own beauties.

Second, the representation graph exists for arbitrary
Lie algebras, not only for $SU(N)$, and the sum path
formula should be straightforwardly generalized
to arbitrary groups, in particular, from the HOMFLY
(for $SU(N)$) to Kauffman (for $SO(N)$) polynomials.

Third, it is now obvious that the Khovanov-Rozhansky approach \cite{KhR}
and the superpolynomials \cite{sup} should have a similar
representations in terms of the multiple sums over paths
(i.e. over the standard Young tableaux).
There are certain advances in this direction for the torus knots
\cite{torus}, particularly close should be the results
by \cite{Negut}.
We emphasize once again that {\bf the sums over Young tableaux
provide the answers for {\it arbitrary} knots},
not only torus: this is a theorem for the HOMFLY polynomials and
a plausible conjecture for the superpolynomials.

\bigskip

To this list one should of course add a massive calculation of the colored
HOMFLY polynomials for non-torus knots, especially in representations
with many lines and rows in the Young diagram, i.e. not just (anti)symmetric representations.
Such examples are crucially important for understanding
the structure of generic knot polynomials, e.g. {\it a la} \cite{IMMMfe,iGFS,fe21,Zreps}
or \cite{MMSle,Anton} or \cite{BEM},
and various relations between them \cite{Gar,MMpols}.

\section*{Acknowledgements}

We are indebted to S.Artamonov, H.Itoyama and A.Popolitov for
fruitful discussions, which helped us to improve the presentation
of the results from \cite{AnoMMM} and \cite{AnoM}
for the purposes of this letter.

Our work is partly supported by Ministry of Education and Science of
the Russian Federation under contract 8207, by the Brazil National Counsel of Scientific and
Technological Development (Al.Mor.), by the Dynasty Foundation (An.Mor.), by NSh-3349.2012.2,
by RFBR grants 13-02-00457 (A.Mir.), 13-02-00478 (Al.Mor.) and 12-02-31078 mol-a (A.A.; An.Mor.),
by joint grant 12-02-92108-Yaf-a, 13-02-90459-Ukr-f-a, 13-02-91371-ST-a.

\end{document}